\documentclass[12pt]{article}

\usepackage[
    geometry=yurie,
    tikz=off,
    biblatex=yurie,
    hyperref=yurie
]{yurie}

\usepackage{amsmath}
\usepackage{amsfonts}
\usepackage{amssymb}
\usepackage{mathrsfs}
\usepackage{mathdots}
\usepackage{mathtools}
\usepackage{mleftright}

\usepackage{xparse}
\usepackage{xstring}
\usepackage{xspace}
\usepackage{xcolor}

\usepackage{tikz}
\usetikzlibrary{arrows.meta, positioning, shapes.geometric}

\numberwithin{equation}{section}

\allowdisplaybreaks

\renewcommand{\geq}{\geqslant}
\renewcommand{\ge}{\geqslant}
\renewcommand{\leq}{\leqslant}

\definecolor{Incoming}{rgb}{0.8,0,0}
\definecolor{Outgoing}{rgb}{0,0,0.8}
\definecolor{Tachyonic}{rgb}{0.5,0.15,0.8}

\newcommand{\inn}{\mathcolor{Incoming}{\mathsf{i}}}
\newcommand{\out}{\mathcolor{Outgoing}{\mathsf{o}}}

\newcommand{\tinyzero}{{\scriptscriptstyle(0)}}

\newcommand{\tinyminusone}{{\scriptscriptstyle(-1)}}

\newcommand{\tinyone}{{\scriptscriptstyle(1)}}

\newcommand{\gluon}{\mathsf{g}}
\newcommand{\graviton}{\mathsf{h}}

\newcommand{\gmode}[2]{\mathsf{G}^{#1}_{#2}}
\newcommand{\hmode}[2]{\mathsf{H}^{#1}_{#2}}

\newcommand{\hhatmode}[2]{\hat{\graviton}^{#1}_{#2}}

\newcommand{\glogmode}[2]{\mathsf{g}^{\tinyzero,#1}_{#2}}
\newcommand{\hlogmode}[2]{\mathsf{h}^{\tinyzero,#1}_{#2}}

\newcommand{\gsoftmode}[2]{\mathsf{g}^{\tinyminusone,#1}_{#2}}
\newcommand{\hsoftmode}[2]{\mathsf{h}^{\tinyminusone,#1}_{#2}}

\newcommand{\ppb}{\bar{\partial}}

\newcommand{\ad}{\mathrm{ad}}

\newcommand{\smallbullet}{\ensuremath{\raisebox{0.25ex}{\tiny$\bullet$}}}

\DeclareFontFamily{U}{jkpmia}{}
\DeclareFontShape{U}{jkpmia}{m}{it}{<->s*jkpmia}{}
\DeclareFontShape{U}{jkpmia}{bx}{it}{<->s*jkpbmia}{}
\DeclareMathAlphabet{\mathfrakalt}{U}{jkpmia}{m}{it}
\SetMathAlphabet{\mathfrakalt}{bold}{U}{jkpmia}{bx}{it}

\newcommand{\winf}{\ensuremath{\mathrm{L}\mathfrakalt{w}_{1+\infty}}\xspace}

\newcommand{\wwedge}{\ensuremath{\mathrm{L}\mathfrakalt{w}_{\wedge}}\xspace}

\ExplSyntaxOn

\cs_new_protected:Npn \yurie_narrow_sign_process:n #1
{
    \tl_set:Nn \l_tmpa_tl {#1}
    \tl_replace_all:Nnn \l_tmpa_tl {+} {\mathord{+}}
    \tl_replace_all:Nnn \l_tmpa_tl {-} {\mathord{-}}
    \tl_use:N \l_tmpa_tl
}

\NewDocumentCommand{\shortForm}{m}{
    \yurie_narrow_sign_process:n {#1}
}

\NewDocumentCommand{\fracShort}{m m}{
    \frac{
        \yurie_narrow_sign_process:n {#1}
    }{
        \yurie_narrow_sign_process:n {#2}
    }
}

\NewDocumentCommand{\gmShort}{m}{
    \Gamma\mleft[ \yurie_narrow_sign_process:n {#1} \mright]
}

\NewDocumentCommand{\btShort}{m}{
    B\mleft( \yurie_narrow_sign_process:n {#1} \mright)
}

\ExplSyntaxOff

\begin{document}

\begin{titlepage}

    \title{Unifying soft and hard dynamics: \\
        The hard current algebra in celestial holography
    }

    \author{Reiko Liu$^{b}$, Wen-Jie Ma$^{a,b}$}

    \date{}

    \maketitle\thispagestyle{empty}

    \address{a}{Fudan Center for Mathematics and Interdisciplinary Study, Fudan University, Shanghai, 200433, China}

    \address{b}{Shanghai Institute for Mathematics and Interdisciplinary Sciences (SIMIS), Shanghai, 200433, China}

    \email{
        reiko.antoneva@foxmail.com,
        wenjie.ma@simis.cn
    }

    \vfill

    \begin{abstract}
        Soft current algebras capture the infrared structure of scattering in asymptotically flat spacetimes, but an analogous algebraic description of finite-energy dynamics has been missing.
        We uncover an infinite-dimensional \emph{hard current algebra} that encodes finite-energy contributions to scattering and implies novel Ward identities. The soft current algebras are not independent but arise naturally from the hard ones.
        This provides a unified algebraic framework underlying quantum theory in flat spacetime.
    \end{abstract}

    \vfill

\end{titlepage}


\begingroup
\hypersetup{linkcolor=black}
\tableofcontents
\endgroup


\clearpage

\section{Introduction}

Symmetry principles have long guided the quest for a unified understanding of the physical world, providing an elegant mathematical language to encode fundamental laws. In quantum field theory and gravity, these principles often manifest as powerful algebraic structures that organize and constrain observables, such as scattering amplitudes.
A particularly fertile arena for such ideas is asymptotically flat spacetime, where the infrared triangle links soft theorems, asymptotic symmetries, and memory effects \cite{Bondi:1962px,Sachs:1962wk,Sachs:1962zza,zel1974radiation,christodoulou1991nonlinear,Barnich:2009se,Barnich:2010eb,Strominger:2013lka,Strominger:2013jfa,He:2014laa,Kapec:2014opa,He:2014cra,Strominger:2014pwa,Kapec:2014opa,He:2015zea,Kapec:2015ena,Campiglia:2015qka,Campiglia:2015kxa,Pasterski:2015tva,Kapec:2016jld,Strominger:2017zoo}.
Celestial holography elevates this picture by proposing a dual conformal field theory (CFT) on the celestial sphere \cite{Cheung:2016iub,Pasterski:2016qvg,Pasterski:2017kqt,Strominger:2017zoo,Raclariu:2021zjz,Pasterski:2021rjz,Pasterski:2021raf,McLoughlin:2022ljp}, where these infrared effects are encoded by infinite-dimensional soft current algebras \cite{Pate:2019mfs,Banerjee:2020zlg,Banerjee:2020vnt,Guevara:2021abz,Himwich:2021dau,Strominger:2021mtt,Banerjee:2021cly,Banerjee:2021dlm,Ball:2021tmb,Adamo:2021lrv,Fan:2022vbz,Banerjee:2023jne,Himwich:2023njb,Banerjee:2023bni,Banerjee:2023zip,Agrawal:2024sju,Ball:2024oqa,Guevara:2025tsm,Banerjee:2025grp,Miller:2025wpq}.
By contrast, the finite-energy (``hard'') dynamics of scattering amplitudes lack an analogous algebraic description.
This raises a central unresolved question:
\begin{quote}
    \itshape
    Is there a single algebraic framework that captures both soft and hard dynamics, and thereby organizes the full S-matrix?
\end{quote}

\begin{figure}[htbp]
    \label{fig:framework}
    \centering

    \begin{tikzpicture}[
            box/.style={draw, thick, rectangle, inner xsep=8pt, inner ysep=8pt, minimum height=0.8cm, align=center, font=\small},
            arrow/.style={-{Stealth[scale=1.0]}, thick, line width=0.8pt},
            label/.style={font=\scriptsize, midway, align=center}
        ]

        \node[box, draw=none, fill=blue!15] (hard-algebra) at (0, 3) {Hard Current Algebra};

        \node[box, draw=none, fill=orange!15] (soft-algebra) at (0, 1) {Soft Current Algebra};

        \node[box, draw=none, fill=blue!15] (hard-sector) at (5, 3) {Hard Scattering};

        \node[box, draw=none, fill=orange!15] (soft-sector) at (5, 1) {Soft Scattering};

        \node[box, draw=none, fill=pink!25] (s-matrix) at (10, 2) {Celestial/Scattering \\ Amplitude};

        \draw[arrow] (hard-algebra) -- (soft-algebra)
        node[label, right] {Conformal Symmetry};

        \draw[arrow] (hard-algebra) -- (hard-sector);

        \draw[arrow] (soft-algebra) -- (soft-sector);

        \draw[arrow] (hard-sector) -- (s-matrix);

        \draw[arrow] (soft-sector) -- (s-matrix);

        \coordinate (hardTop) at ([shift={(0,1)}]hard-algebra.north);

        \coordinate (smTop) at (s-matrix.north |- hardTop);

        \draw[arrow] (hard-algebra.north) -- (hardTop)
        -- node[label, midway, below] {New Ward Identity} (smTop)
        -- (s-matrix.north);

    \end{tikzpicture}

    \caption{The unified algebraic framework}

\end{figure}
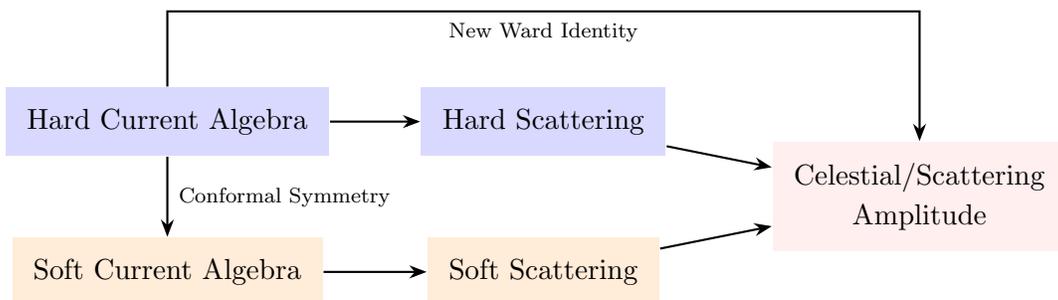

We address this question by uncovering a new infinite-dimensional algebraic structure --- the \emph{hard current algebra}.
This algebra is constructed from subleading Laurent coefficients of celestial primaries at integer conformal dimensions.
In the bulk, its generators capture finite-energy contributions to scattering amplitudes; on the boundary, they imply novel Ward identities that constrain celestial amplitudes beyond the soft sector.
We further show that hard and soft currents furnish logarithmic representations of the conformal group, and the known soft current algebras are not independent but arise naturally from the hard ones.
The detailed realization of the hard current algebra depends on the theory through collinear operator product expansion (OPE), but the framework is universal and provides a unified algebraic structure for the S-matrix, interpolating between the soft and hard sectors  (see Figure \ref{fig:framework}).

\section{Hard current algebra}\label{sec:hard current algebra}

Starting from the OPEs of gravitons and gluons, we construct the hard current algebras that extend the soft current algebras. We only consider positive-helicity outgoing particles.

\textbf{Graviton.}
At tree level, the collinear OPE of gravitons is \cite{Fan:2019emx,Pate:2019lpp}
\begin{equation}
    \label{eq:graviton_OPE}
    \graviton_{\Delta_1}\graviton_{\Delta_2}
    \sim
    \sum_{t=0}^{\infty}
    \frac{\bar{z}_{12}^{t+1}}{z_{12}}
    \frac{1}{t!}
    \btShort{\Delta_1+t-1,\Delta_2-1}
    \bar{\partial}^t_2
    \graviton_{\Delta_1+\Delta_2}
    ,
\end{equation}
where $z_{12}\equiv z_1 - z_2$ and $\bar{\partial}_i\equiv\partial_{\bar{z}_i}$. We display only anti-holomorphic descendants, which carry the $1/z_{12}$ pole responsible for the commutators below. The graviton operator has simple poles at $\Delta = k$ for integer $k \leq 2$, with Laurent expansion $\graviton_{\Delta}(z,\bar{z})=\sum_{i=-1}^{\infty} (\Delta - k)^i\graviton^{(i)}_{k}(z,\bar{z})$. The conformally soft gravitons $\graviton^{\tinyminusone}_{k}$ are primaries with weights $(h,\bar{h})=\bigl(\frac{k+2}{2},\frac{k-2}{2}\bigr)$ that satisfy
\begin{equation}
    \label{eq:soft_graviton_null_state}
    \ppb^{3-k}\graviton^{\tinyminusone}_{k}(z,\bar{z}) = 0.
\end{equation}
They generate (the wedge subalgebra of) the \winf algebra \cite{Strominger:2021mtt}.

\begin{figure}[htbp]
    \label{fig:hard current algebra in w-algebra}
    \centering

    \tikzset{every picture/.style={line width=0.75pt}} 
    \scalebox{0.8}{
        \begin{tikzpicture}[x=0.75pt,y=0.75pt,yscale=-1,xscale=1]

            \draw  [color={rgb, 255:red, 255; green, 255; blue, 255 }  ,draw opacity=1 ][fill={rgb, 255:red, 80; green, 227; blue, 194 }  ,fill opacity=0.2 ] (240,240) -- (0,0) -- (480,0) -- cycle ;
            \draw [color={rgb, 255:red, 155; green, 155; blue, 155 }  ,draw opacity=1 ] [dash pattern={on 4.5pt off 4.5pt}]  (240,240) -- (240,3) ;
            \draw [shift={(240,0)}, rotate = 90] [fill={rgb, 255:red, 155; green, 155; blue, 155 }  ,fill opacity=1 ][line width=0.08]  [draw opacity=0] (8.93,-4.29) -- (0,0) -- (8.93,4.29) -- cycle    ;
            \draw [color={rgb, 255:red, 155; green, 155; blue, 155 }  ,draw opacity=1 ] [dash pattern={on 4.5pt off 4.5pt}]  (0,240) -- (477,240) ;
            \draw [shift={(480,240)}, rotate = 180] [fill={rgb, 255:red, 155; green, 155; blue, 155 }  ,fill opacity=1 ][line width=0.08]  [draw opacity=0] (8.93,-4.29) -- (0,0) -- (8.93,4.29) -- cycle    ;
            \draw  [draw opacity=0][fill={rgb, 255:red, 208; green, 2; blue, 27 }  ,fill opacity=0.2 ] (200,240) -- (0,40) -- (0,240) -- cycle ;
            \draw  [draw opacity=0][fill={rgb, 255:red, 74; green, 144; blue, 226 }  ,fill opacity=0.2 ] (280,240) -- (480,40) -- (480,240) -- cycle ;
            \draw  [fill={rgb, 255:red, 0; green, 0; blue, 0 }  ,fill opacity=1 ] (240,241) .. controls (239.45,241) and (239,240.55) .. (239,240) .. controls (239,239.45) and (239.45,239) .. (240,239) .. controls (240.55,239) and (241,239.45) .. (241,240) .. controls (241,240.55) and (240.55,241) .. (240,241) -- cycle ;
            \draw  [fill={rgb, 255:red, 0; green, 0; blue, 0 }  ,fill opacity=1 ] (200,241) .. controls (199.45,241) and (199,240.55) .. (199,240) .. controls (199,239.45) and (199.45,239) .. (200,239) .. controls (200.55,239) and (201,239.45) .. (201,240) .. controls (201,240.55) and (200.55,241) .. (200,241) -- cycle ;
            \draw  [fill={rgb, 255:red, 0; green, 0; blue, 0 }  ,fill opacity=1 ] (280,241) .. controls (279.45,241) and (279,240.55) .. (279,240) .. controls (279,239.45) and (279.45,239) .. (280,239) .. controls (280.55,239) and (281,239.45) .. (281,240) .. controls (281,240.55) and (280.55,241) .. (280,241) -- cycle ;

            \draw (250,10) node [anchor=north west][inner sep=0.75pt]    {$p$};
            \draw (470,230) node [anchor=south east][inner sep=0.75pt]    {$n'$};
            \draw (360,180) node [anchor=north west][inner sep=0.75pt]    {$\hmode{k}{m,n} ,\ n\geq 3-k$};
            \draw (35,180) node [anchor=north west][inner sep=0.75pt]    {$\hmode{k}{m,n} ,\ n\leq -1$};
            \draw (222,120) node [anchor=north west][inner sep=0.75pt]    {\scalebox{1.2}{\wwedge}};
            \draw (225,220) node [anchor=north west][inner sep=0.75pt]    {\scalebox{0.8}{$(0,1)$}};
            \draw (175,220) node [anchor=north west][inner sep=0.75pt]    {\scalebox{0.8}{$( -1,1)$}};
            \draw (265,220) node [anchor=north west][inner sep=0.75pt]    {\scalebox{0.8}{$( 1,1)$}};

        \end{tikzpicture}
    }

    \caption{
        Relation between the hard current algebra and the \winf algebra.
        Each integer point $(n,p')$ denotes a generator $w_{m,n'}^{p} = \hmode{2-k/2}{m,n-1+k/2}$ of the full \winf algebra.
        The green triangle represents the wedge subalgebra \wwedge.
        The blue/red triangle corresponds to the first/second line of the hard current algebra \eqref{eq:graviton hard current algebra} in sector (I).
    }

\end{figure}
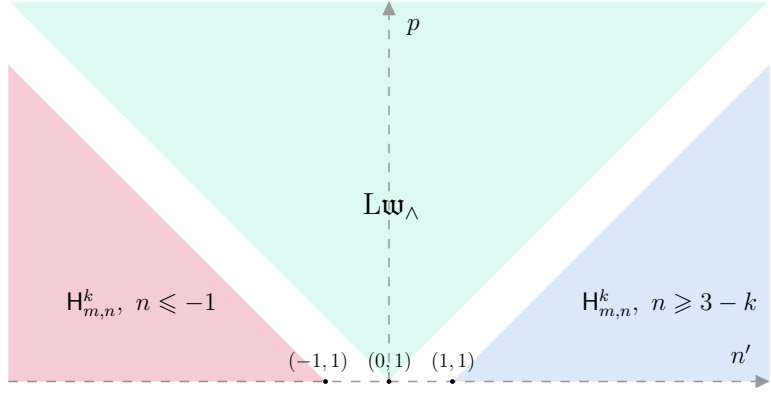

Our focus is the algebraic structure generated by the subleading operator $\graviton^{\tinyzero}_{k}$.
They are not closed under the OPE, \eg, $\graviton^{\tinyone}_{0}\in \graviton^{\tinyzero}_{0}\graviton^{\tinyzero}_{0}$. However, acting with sufficiently many $\bar{\partial}$ derivatives removes all contributions from $\graviton^{(i)}_{k_1+k_2}$ with $i\neq 0$. For $k_1,k_2\leq 2$ and $s_1\geq 3-k_1$, $s_2\geq 3-k_2$, \eqref{eq:graviton_OPE} implies:\footnote{Appendix \ref{app:OPE subleading operator} contains a detailed derivation.}
\begin{align}\label{eq:graviton OPE subleading operator}
    \bar{\partial}^{s_1}_1\graviton^{\tinyzero}_{k_1}\bar{\partial}_2^{s_2}\graviton^{\tinyzero}_{k_2}
    \sim
    \sum_{t=0}^{\infty}
    \frac{\bar{z}_{12}^{t}}{z_{12}}
    \fracShort{s_2(2-k_1)-(2-k_2)(t+s_1)}{t!(s_2+k_2-2)}
    \btShort{s_1+k_1+t,s_2+k_2-1}
    \bar{\partial}_2^{t+s_1+s_2-1}\graviton^{\tinyzero}_{k_1+k_2}.
\end{align}
There are two closed subsectors in this OPE: (I) $k=2,0,-2,\dots$ and (II) $k=0,-1,-2,\dots$.
Expanding in the modes,\footnote{Appendix \ref{app:soft algebra from hard algebra} contains a detailed derivation.}
\begin{align}\label{eq:graviton mode expansion}
    \bar{\partial}^{3-k}\graviton^{\tinyzero}_{k}(z,\bar{z})=\sum_{m,n\in\mathbb{Z}}z^{-m-1}\bar{z}^{-n-1}\hmode{k}{m,n},
\end{align}
yields the commutator:\footnote{Appendix \ref{app:hard current algebra} contains a detailed derivation.}
\begin{align}\label{eq:graviton hard current algebra}
    [\hmode{k_1}{m_1,n_1},\hmode{k_2}{m_2,n_2}]
    &=
    \begin{cases}
        [n_2(2-k_1)-n_1(2-k_2)]\hmode{k_1+k_2}{m_1+m_2,n_1+n_2-1}& \text{for } n_{1}\geq 3-k_{1}, n_{2}\geq 3-k_{2},
        \\
        [n_1(2-k_2)-n_2(2-k_1)]\hmode{k_1+k_2}{m_1+m_2,n_1+n_2-1}& \text{for } n_1,n_2\leq-1.
    \end{cases}
\end{align}
Although the full set $\{\hmode{k}{m,n}\}$ with $k\leq 2$ is not closed, each subsector (I) and (II) forms a closed \textit{hard current algebra}. In sector (I), this algebra is related to the \winf algebra; see Figure \ref{fig:hard current algebra in w-algebra}.

\textbf{Gluon.}
The tree-level collinear OPE of gluons is \cite{Fan:2019emx,Pate:2019lpp}
\begin{equation}
    \label{eq: gluon OPE}
    \gluon_{\Delta_1}^{a}\gluon^{b}_{\Delta_2}
    \sim
    -i f^{abc}
    \sum_{t=0}^{\infty}
    \frac{\bar{z}_{12}^{t}}{z_{12}}
    \fracShort{1}{t!}
    \btShort{\Delta_{1}+t-1,\Delta_{2}-1}
    \bar{\partial}_2^{t}
    \gluon^{c}_{\Delta_{1}+\Delta_{2}-1}
    .
\end{equation}
Gluon operators exhibit simple poles at $\Delta=k$ for integer $k \leq 1$, with Laurent expansion $\gluon^{a}_{\Delta}(z,\bar{z}) = \sum_{i=-1}^{\infty} (\Delta - k)^i\gluon^{(i),a}_{k}(z,\bar{z})$.
The conformally soft primaries $\gluon^{\tinyminusone}_{k}$ with weights $(h,\bar{h}) = \bigl( \frac{k+1}{2}, \frac{k-1}{2} \bigr)$ satisfy
\begin{equation}
    \label{eq:soft gluon null state}
    \ppb^{2-k}\gluon^{\tinyminusone,a}_{k}(z,\bar{z}) = 0,
\end{equation}
and generate the soft gluon algebra.

Subleading operators $\bar{\partial}^{s}\gluon^{\tinyzero}_{k}$ in the sector $s \geq 2-k$ satisfy the OPE:\footnote{Appendix \ref{app:OPE subleading operator} contains a detailed derivation.}
\begin{align}\label{eq:gluon OPE subleading operator}
    &\bar{\partial}^{s_1}_1\gluon^{\tinyzero,a}_{k_1}\bar{\partial}_2^{s_2}\gluon^{\tinyzero,b}_{k_2}\sim-i f^{abc}\sum_{t=0}^{\infty}\frac{\bar{z}_{12}^{t}}{z_{12}}\frac{B(s_1+k_1+t-1,s_2+k_2-1)}{t!}\bar{\partial}_2^{s_1+s_2+t}\gluon^{\tinyzero,c}_{k_1+k_2-1}.
\end{align}
Expanding in the modes,
\begin{align}\label{eq:gluon mode expansion}
    \bar{\partial}^{2-k}\gluon^{\tinyzero,a}_{k}(z,\bar{z})=\sum_{m,n\in\mathbb{Z}}z^{-m-1}\bar{z}^{-n-1}\gmode{m,n}{k,a},
\end{align}
yields the hard current algebra:\footnote{Appendix \ref{app:hard current algebra} contains a detailed derivation.}
\begin{align}\label{eq:gluon hard current algebra}
    [\gmode{m_1,n_1}{k_1,a},\gmode{m_2,n_2}{k_2,b}]
    &=
    \begin{cases}
        i f^{abc}\gmode{m_1+m_2,n_1+n_2}{k_1+k_2-1,c}        & \text{for } n_1\geq 2-k_1, n_2\geq 2-k_2,
        \\
        -i f^{abc}\gmode{m_1+m_2,n_1+n_2}{k_1+k_2-1,c}& \text{for } n_1,n_2\leq-1,
    \end{cases}
\end{align}
which closes for $k \leq 1$.

\section{Relation to finite-energy scattering}

We now show that the hard current algebra encodes the finite-energy (hard) part of scattering amplitudes.
On the boundary, the gluon operator admits a Mittag–Leffler expansion in $\Delta$ as
\begin{align}
    \gluon^{a}_{\Delta}(z,\bar{z})
    =
    \sum_{k'\leq 1}
    \frac{\gluon^{\tinyminusone,a}_{k'}(z,\bar{z})}{\Delta-k'}+R_{\Delta}^a(z,\bar{z}),
\end{align}
where $R_{\Delta}^a$ is holomorphic in $\Delta$. Comparing with the Laurent expansion in $\Delta$ and using the current conservation, we have
\begin{align}
    \bar{\partial}^{2-k}\gluon^{\tinyzero,a}_{k}
    =
    \sum_{k'\leq k-1}
    \frac{\bar{\partial}^{2-k}\gluon^{\tinyminusone,a}_{k'}}{k-k'}+\bar{\partial}^{2-k}R^a_{k}.
\end{align}
Thus the generators $\gluon^{\tinyzero,a}_{k}$ package the soft data together with the additional physical information contained in $R_{\Delta}^a$.

In the bulk, the operator $R_{\Delta}^a$ carries the hard-sector information of scattering amplitudes.
To see this, we consider the celestial amplitude,
\begin{align}
    \mathcal{A}_{\Delta}(z,\cdots)=\int_0^{\infty}d\omega\omega^{\Delta-1}\mathcal{T}(q,\cdots),
\end{align}
where only the first gluon and its Mellin transform are shown. Here we assume that the scattering amplitude decays sufficiently fast in the UV.
We separate the $\omega$-integral into the soft region $\omega\in [0,1]$ and the hard region $\omega\in [1,\infty)$. In the soft region, the scattering amplitude can be expanded using the soft theorem in powers of $\omega^{-k}$ with integer $k\leq 1$. Performing the $\omega$-integral yields a term proportional to $(\Delta-k)^{-1}$, corresponding to the soft gluon $\gluon^{\tinyminusone,a}_{k}$. In contrast, the hard-region integral is regular and corresponds to the operator $R_{\Delta}^a$ on the celestial sphere.

This discussion confirms that the algebras \eqref{eq:graviton hard current algebra} and \eqref{eq:gluon hard current algebra} indeed capture the hard contributions, justifying their designation as hard current algebras.

\section{Ward identity}
\label{sec:ward identities}

In standard CFT, current algebras act on correlators via Ward identities, which in the soft case reproduce soft theorems of scattering amplitudes.
To establish the hard current algebra as a bona fide current algebra, we must demonstrate that the associated hard charges act consistently on primary operators and that correlators obey the corresponding Ward identities.
In what follows we derive these Ward identities and verify them explicitly for three-point regular celestial amplitudes \cite{Liu:2025voe}.

We begin with gluons and consider the radially ordered correlator
\begin{align}
    \lim_{z,\bar{z}\to 0}\langle0|\mathcal{R}\{\mathcal{O}_1\cdots\mathcal{O}_N\}\gluon_{k}^{\tinyzero,a}(z,\bar{z})|0\rangle
    .
\end{align}
Here all other insertions are away from $0$ and $\infty$, and it is natural to assume that this correlator is regular for $z,\bar{z}\sim 0$ and $\infty$. Then the mode expansion \eqref{eq:log mode expansion} together with \eqref{eq:hard mode and logmode} implies
\begin{align}\label{eq:hmode vacuum=0}
    \begin{split}
        &\gmode{k,a}{m,n}|0\rangle=0,
        \quad
        m\geq0 \text{ and } n\geq1,\\
        &\langle0|\gmode{k,a}{m,n}=0,
        \quad
        m\leq k-1 \text{ and } n\leq-1,
    \end{split}
\end{align}
where the second identity follows from the first by inversion. Hence $\langle\gmode{k,a}{m,n}\mathcal{O}_1\cdots\mathcal{O}_N\rangle=0$ for $m\leq k-1$ and $n\leq-1$. Then using $\gmode{k,a}{m,n}|0\rangle=\partial^{-m-1}\bar{\partial}^{1-k-n}\gluon^{\tinyzero,a}_{k}(0)$ for $m\leq-1$ and $n\leq1-k$, we obtain
\begin{align}\label{eq:gluon ward identity}
    \partial^{-m-1}\bar{\partial}^{1-k-n}\langle\gluon^{\tinyzero,a}_{k}(0)\mathcal{O}_1\cdots \mathcal{O}_N\rangle=-\Gamma[-m]\Gamma[-n]\sum_{i=1}^{N}\langle\mathcal{O}_1\cdots[\gmode{m,n}{k,a},\mathcal{O}_i]\cdots\mathcal{O}_N\rangle,
\end{align}
valid for $m\leq\text{min}(k-1,-1)$ and $n\leq-1$.
The graviton case is analogous and yields
\begin{align}\label{eq:graviton ward identity}
    \partial^{-m-1}\bar{\partial}^{2-k-n}\langle\graviton^{\tinyzero}_{k}(0)\mathcal{O}_1\cdots \mathcal{O}_N\rangle=-\Gamma[-m]\Gamma[-n]\sum_{i=1}^{N}\langle\mathcal{O}_1\cdots[\hmode{k}{m,n},\mathcal{O}_i]\cdots\mathcal{O}_N\rangle,
\end{align}
valid for $m\leq\text{min}(k,-1)$ and $n\leq-1$.

To verify the Ward identity \eqref{eq:gluon ward identity}, we derive the action of hard charges for $n\leq-1$ on gluons,
\begin{align}\label{eq:gmode on g}
    &[\gmode{\out,k,a}{m,n},\gluon^{\out,b}_{\Delta,+}(z,\bar{z})]
    =
    -i f^{abc}\sum_{t=1-k-n}^{\infty}\binom{t+k-2}{-1-n}\frac{z^{m}(-\bar{z})^{t+k+n-1}}{(\Delta-1)_{t+k-1}}\bar{\partial}^t\gluon^{\out,c}_{\Delta+k-1,+}(z,\bar{z}),
    \\
    \label{eq:gmode on shadow g}
    &[\gmode{\out,k,a}{m,n},\widetilde{\gluon}^{\inn,b}_{\Delta,+}(z,\bar{z})]
    =
    -i f^{abc}\delta_{k,1}z^m\bar{z}^{n}\widetilde{\gluon}^{\inn,c}_{\Delta,+}(z,\bar{z}).
\end{align}
Here $\widetilde{\gluon}$ is the shadow gluon operator and $\out$/$\inn$ denote outgoing/incoming. These actions follow from the OPEs $\gluon\gluon\sim\gluon$ and $\gluon\widetilde{\gluon}\sim\widetilde{\gluon}$ \cite{Himwich:2025bza}, and furnish a representation of the hard current algebra for $n\leq-1$.

Now we verify \eqref{eq:gluon ward identity} on the three-point regular celestial amplitude,
\begin{equation}
    \label{eq:gluon ward identity three-point}
    -\fracShort{\partial^{-m-1}\bar{\partial}^{1-k-n}}{(-m-1)!(-n-1)!}
    \langle\gluon^{\out,\tinyzero,a}_{k,+}(0)\gluon^{\out,b}_{\Delta_2,+}\widetilde{\gluon}^{\inn,c}_{\Delta_3,+}\rangle
    =
    \langle [\gmode{k,a}{m,n},\gluon^{\out,b}_{\Delta_2,+}] \widetilde{\gluon}^{\inn,c}_{\Delta_3,+}\rangle
    +
    \langle \gluon^{\out,b}_{\Delta_2,+} [\gmode{k,a}{m,n},\widetilde{\gluon}^{\inn,c}_{\Delta_3,+}] \rangle
    .
\end{equation}
Here we insert a shadow gluon so that the two-point function $\langle \gluon^{\out,a}_{\Delta,+}\widetilde{\gluon}^{\inn,b}_{\Delta,+} \rangle$ takes the standard CFT form.
To evaluate the left-hand side, we use the regular celestial amplitude $\langle\gluon^{\out,a}_{\Delta_{1},+}\gluon^{\out,b}_{\Delta_2,+}\widetilde{\gluon}^{\inn,c}_{\Delta_3,+}\rangle$. Its coordinate dependence takes the standard form, and the three-point coefficient factorizes into the OPE coefficient $\btShort{\Delta_1-1,\Delta_2-1}$ and a two-point coefficient $C_{2}$ \cite{Liu:2025voe}. Using $\bar{\partial}^{s}\gluon^{\out,\tinyzero,a}_{k,+} = \bar{\partial}^{s}\gluon^{\out,a}_{\Delta,+}\big|_{\Delta=k}$ for $s\geq 2-k$, we find
\begin{align}\label{eq:gluon 3pt}
    \textrm{l.h.s.}
    =
    if^{abc}
    \shortForm{(-m-1)!(-n-1)!(2-\Delta_2)_{1-k}}
    \delta_{\Delta_2+k-1,\Delta_3}C_2
    (z_2^m-\delta_{n,0}z_3^m)
    \bar{z}_{3}^{n}z_{23}^{-k-\Delta_2}\bar{z}_{23}^{1-\Delta_2}
    .
\end{align}
Evaluating the right-hand side by applying \eqref{eq:gmode on g} and \eqref{eq:gmode on shadow g} to \eqref{eq:gluon ward identity three-point} reproduces exactly the expression in \eqref{eq:gluon 3pt}.

\section{From hard to soft current algebra}

A central result is that the soft current algebra is not an independent structure but emerges from the more fundamental hard current algebra in \eqref{eq:graviton hard current algebra} and \eqref{eq:gluon hard current algebra}.
This follows from the observation that the soft current and its associated hard operator form a logarithmic pair under $\mathrm{SL}(2,\mathbb{C})$; for the graviton,
\begin{align}
    \label{eq:conformal algebra subleading graviton}
    [\bar{L}_{n},\graviton^{\tinyzero}_{k}(z,\bar{z})]
    &=
    \bigl(\bar{z}^{n+1}\bar{\partial}+\tfrac{1}{2}(k-2)(n+1)\bar{z}^n\bigr)\graviton^{\tinyzero}_{k}(z,\bar{z})
    +\tfrac{1}{2}(n+1)\bar{z}^n\graviton^{\tinyminusone}_{k}(z,\bar{z})
    ,
\end{align}
which implies that soft operators are generated by acting with conformal generators on hard operators. We make this explicit via mode expansions. The soft currents have the truncated expansions
\begin{align}
    \label{eq:soft mode expansion}
    \begin{split}
        &\graviton^{\tinyminusone}_k(z,\bar{z})=\sum_{m\in\mathbb{Z}}\sum_{k-2\leq n \leq 0}z^{-m-1}\bar{z}^{-n}\hsoftmode{k}{m,n},\\
        &\gluon^{\tinyminusone}_k(z,\bar{z})=\sum_{m\in\mathbb{Z}}\sum_{k-1\leq n \leq 0}z^{-m-1}\bar{z}^{-n}\gsoftmode{k,a}{m,n}
        ,
    \end{split}
\end{align}
while the hard operators are expanded as
\begin{align}
    \label{eq:log mode expansion}
    \begin{split}
        &\graviton^{\tinyzero}_{k}(z,\bar{z})= \sum_{m,n\in\mathbb{Z}}z^{-m-1}\bar{z}^{-n-k+2}\hlogmode{k}{m,n},\\
        &\gluon^{\tinyzero,a}_{k}(z,\bar{z})=\sum_{m,n\in\mathbb{Z}}z^{-m-1}\bar{z}^{-n-k+1}\glogmode{k,a}{m,n}.
    \end{split}
\end{align}
Matching \eqref{eq:graviton mode expansion} and \eqref{eq:gluon mode expansion} with \eqref{eq:log mode expansion} gives
\begin{align}\label{eq:hard mode and logmode}
    \begin{split}
        &\hmode{k}{m,n}=(-n)_{3-k}\hlogmode{k}{m,n},\;\;\gmode{m,n}{k,a}=(-n)_{2-k}\glogmode{k,a}{m,n}.
    \end{split}
\end{align}
In particular, $\hmode{k}{m,n}=0$ for $0\leq n\leq 2-k$ and $\gmode{m,n}{k,a}=0$ for $0\leq n\leq 1-k$. Using \eqref{eq:conformal algebra subleading graviton} then yields the generating relations
\begin{align}\label{eq:Lbar1 on hmode to hsoftmode}
    \begin{split}
        &
        \ad^i_{\bar{L}_1}
        \hmode{k}{m,-1}
        =(-1)^{i+1}(3-k)!(i-1)!\hsoftmode{k}{m,k+i-3},\\
        &\ad_{\bar{L}_1}^i\gmode{m,-1}{k,a}=(-1)^{i+1}(2-k)!(i-1)!\gsoftmode{k,a}{m,k+i-2},
    \end{split}
\end{align}
for $i\ge 1$, where $\ad^i_{\bar{L}_1}(\smallbullet)\equiv[\bar{L}_1,\cdots,[\bar{L}_1,\smallbullet]]$.

Equation \eqref{eq:Lbar1 on hmode to hsoftmode} shows that repeated adjoint action of $\bar{L}_1$ on a single hard mode generates the full tower of soft modes. The soft current algebra is therefore fixed by the hard current algebra.

\textbf{Reconstruction.} We illustrate the reconstruction of the soft gluon algebra. The commutators $[\gsoftmode{k_1,a}{s_1,n_1},\gsoftmode{k_2,b}{s_2,n_2}]$ with $n_1+n_2=-n_0$ and $0 \leq n_0 \leq 2-k_1-k_2$ are determined by acting with
\begin{align}\label{eq:equations to solve soft gluon algebra}
    \ad_{\bar{L}_{-1}}^{n_0-i}\ad^{4-k_1-k_2-i}_{\bar{L}_1}[\gmode{s_1,-1}{k_1,a},\gmode{s_2,-1}{k_2,b}],\;\;0\leq i\leq n_0.
\end{align}
This produces $n_0+1$ independent equations, which uniquely fix the $n_0+1$ unknown commutators.

Consider the case $n_0=0$. Applying $\ad^{4-k_1-k_2}_{\bar{L}_{1}}$ to $[\gmode{m_1,-1}{k_1,a},\gmode{m_2,-1}{k_2,b}]$ and using the Leibniz rule with \eqref{eq:Lbar1 on hmode to hsoftmode} gives
\begin{align}
    \ad^{4-k_1-k_2}_{\bar{L}_1}[\gmode{m_1,-1}{k_1,a},\gmode{m_2,-1}{k_2,b}]
    &=(-1)^{k_1+k_2}(4-k_1-k_2)!(1-k_1)!(1-k_2)![\gsoftmode{k_1,a}{m_1,0},\gsoftmode{k_2,b}{m_2,0}].
\end{align}
The left-hand side can be evaluated independently using the hard current algebra in \eqref{eq:gluon hard current algebra}, leading to
\begin{align}
    \ad^{4-k_1-k_2}_{\bar{L}_1}[\gmode{m_1,-1}{k_1,a},\gmode{m_2,-1}{k_2,b}]
    &=-if^{abc}(-1)^{k_1+k_2}(4-k_1-k_2)!(2-k_1-k_2)!\gsoftmode{k_1+k_2-1,c}{m_1+m_2,0}.
\end{align}
Equating the two expressions reproduces the commutator $[\gsoftmode{k_1}{m_1,0},\gsoftmode{k_2}{m_2,0}]$ in \cite{Guevara:2021abz}.

The remaining commutators follow analogously by solving the equations generated by \eqref{eq:equations to solve soft gluon algebra}. In this sense, the full soft current algebra is encoded in the hard current algebra.



\section{Future directions}

There are several future directions.
First, our work suggests a natural extension of the infrared triangle to a more fundamental ``soft-hard quadrangle'' with the hard current algebra as the fourth vertex. It would be interesting to explore this structure in detail, examining its implications on asymptotic symmetries and memory effects.
Second, the logarithmic representations indicate that celestial CFTs lie in the class of logarithmic CFTs, see also \cite{Fiorucci:2023lpb,Bissi:2024brf}. This provides a concrete and physical motivation to study such non-unitary representations.
Third, it is also important to extend the tree-level, collinear-OPE derivation of the hard current algebra to loop level and to other theories.
Finally, the Ward identities impose novel constraints on celestial amplitudes. Verifying them at higher points and exploring their implications on scattering amplitudes are natural next steps. Combining these constraints with the celestial optical theorem \cite{Liu:2024vmx} and the amplitude-reconstruction method \cite{Liu:2025dhh} may enable a bootstrap program.


\section*{Acknowledgements}

The authors would like to thank Hamed Adami and Yehao Zhou for useful discussions.
WJM is supported by the National Natural Science Foundation of China No. 12405082 and Shanghai Pujiang Program No. 24PJA118.


\clearpage

\appendix

\section{Derivation of OPEs for subleading operators}\label{app:OPE subleading operator}
In this appendix, we derive the OPEs for the subleading operators $\bar{\partial}^s\graviton^{\tinyzero}_k$ and $\bar{\partial}^s\gluon^{\tinyzero}_k$.

\textbf{Gluon.} We begin with the gluon case. Substituting the Laurent expansion into the gluon OPE and applying the differential operators $\bar{\partial}^{s_1}_1$ and $\bar{\partial}^{s_2}_2$ with $s_1\geq 2-k_1$ and $s_2\geq 2-k_2$ to both sides gives
\begin{align}
    &\sum_{i=-1}^{\infty}(\Delta_1-k_1)^i\bar{\partial}^{s_1}_1\gluon^{(i),a}_{k_1,+}\sum_{j=-1}^{\infty}(\Delta_2-k_2)^j\bar{\partial}_2^{s_2}\gluon^{(j),b}_{k_2}\sim-i f^{abc}\sum_{r'=0}^{s_2}\sum_{r=s_1+s_2-r'}^{\infty}\frac{(-1)^{s_2}}{r'!(r+r'-s_1-s_2)!}\\\nonumber
    &\qquad\times\frac{\bar{z}_{12}^{r+r'-s_1-s_2}}{z_{12}}B(\Delta_1-1+r,\Delta_2-1)(-s_2)_{r'}\sum_{l=-1}^{\infty}(\Delta_{1}+\Delta_2-k_1-k_2)^l\bar{\partial}_2^{r+r'}\gluon^{(l),c}_{k_1+k_2-1}.
\end{align}
Changing summation variables from $(r,r')$ to $(r,t=r+r')$, employing the conservation condition of the conformally soft operators, and summing over $r'$ yields
\begin{align}
    &\sum_{i=0}^{\infty}(\Delta_1-k_1)^i\bar{\partial}^{s_1}_1\gluon^{(i),a}_{k_1}\sum_{j=0}^{\infty}(\Delta_2-k_2)^j\bar{\partial}_2^{s_2}\gluon^{(j),b}_{k_2}\sim-i f^{abc}\frac{B(\Delta_1-1+t,\Delta_2-1)(\Delta_2-1)_{s_2}}{(2-t-\Delta_1)_{s_2}}\\\nonumber
    &\qquad\qquad\times\sum_{t=s_1+s_2}^{\infty}\frac{(-1)^{s_2}}{(t-s_1-s_2)!}\frac{\bar{z}_{12}^{t-s_1-s_2}}{z_{12}}\sum_{l=0}^{\infty}(\Delta_{1}+\Delta_2-k_1-k_2)^l\bar{\partial}_2^{t}\gluon^{(l),c}_{k_1+k_2-1}.
\end{align}
We now expand both sides of this OPE around $\Delta_1=k_1$ and $\Delta_2=k_2$ to order $(\Delta_1-k_1)^0(\Delta_2-k_2)^0$, obtaining
\begin{align}
    \bar{\partial}^{s_1}_1\gluon^{(i),a}_{k_1}\bar{\partial}_2^{s_2}\gluon^{(j),b}_{k_2}\sim-i f^{abc}\sum_{t=s_1+s_2}^{\infty}\frac{\bar{z}_{12}^{t-s_1-s_2}}{z_{12}}\frac{(s_2+k_2-2)!}{(t+k_1-s_2-1)_{s_2+k_2-1}(t-s_1-s_2)!}\bar{\partial}_2^{t}\gluon^{\tinyzero,c}_{k_1+k_2-1}.
\end{align}
Finally, shifting the summation index $t\rightarrow t+s_1+s_2$ produces the OPE presented in the main text.

\textbf{Graviton.} An analogous derivation applies to the subleading graviton operator. Substituting the Laurent expansion into the graviton OPE and acting with the differential operators $\bar{\partial}^{s_1}_1$ and $\bar{\partial}^{s_2}_2$, where $s_1\geq 3-k_1$ and $s_2\geq 3-k_2$, on both sides produces
\begin{align}
    \nonumber&\sum_{i=-1}^{\infty}(\Delta_1-k_1)^i\bar{\partial}^{s_1}_1\graviton^{(i)}_{k_1}\sum_{j=-1}^{\infty}(\Delta_2-k_2)^j\bar{\partial}_2^{s_2}\graviton^{(j)}_{k_2}\sim\sum_{r'=0}^{s_2}\frac{(t-r'+1)B(\Delta_1-1+t-r',\Delta_2-1)(-s_2)_{r'}}{r'!}\\
    &\qquad\times\sum_{t=s_1+s_2-1}^{\infty}\frac{(-1)^{s_2}}{(t+1-s_1-s_2)!}\frac{\bar{z}_{12}^{t+1-s_1-s_2}}{z_{12}}\sum_{l=-1}^{\infty}(\Delta_{1}+\Delta_2-k_1-k_2)^l\bar{\partial}_2^{t}\graviton^{(l)}_{k_1+k_2}.
\end{align}
Imposing the conservation condition for the conformally soft operators, performing the sum over $r'$, and expanding both sides around $\Delta_1=k_1$ and $\Delta_2=k_2$ to order $(\Delta_1-k_1)^0(\Delta_2-k_2)^0$ yields the graviton OPE found in the main text.

\section{Hard current algebra}\label{app:hard current algebra}
This appendix details the derivation of the hard current algebra for gluons. The derivation for gravitons follows analogously and is omitted here for brevity. Starting from the mode expansion
\begin{align}
    \bar{\partial}^{2-k}\gluon^{\tinyzero,a}_{k}(z,\bar{z})=\sum_{n\in\mathbb{Z}}\bar{z}^{-n-1}\gmode{n}{k,a}(z),
\end{align}
the modes are extracted via the contour integral
\begin{align}
    \gmode{n}{k,a}(z)=\oint\frac{d\bar{z}}{2\pi i}\bar{z}^{n}\bar{\partial}^{2-k}\gluon^{\tinyzero,a}_{k}(z,\bar{z}).
\end{align}
The algebra generated by these modes follows from computing the triple contour integral
\begin{align}
    [\gmode{n_1}{k_1,a},\gmode{n_2}{k_2,b}](z_2)=\oint_{|\bar{z}_1|<\epsilon}\frac{d\bar{z}_1}{2\pi i}\bar{z}_1^{n_1}\oint_{|\bar{z}_2|<\epsilon}\frac{d\bar{z}_2}{2\pi i}\bar{z}_2^{n_2}\oint_{|z_{12}|<\epsilon}\bar{\partial}^{2-k_1}\gluon^{\tinyzero,a}_{k_1}(z_1,\bar{z}_1)\bar{\partial}^{2-k_2}\gluon^{\tinyzero,b}_{k_2}(z_2,\bar{z}_2).
\end{align}
Substituting the OPE for the subleading gluon operator yields
\begin{align}
    [\gmode{n_1}{k_1,a},\gmode{n_2}{k_2,b}](z_2)&=\sum_{t=0}^{\infty}\frac{-i f^{abc}}{(t+1)!}\oint_{|\bar{z}_1|<\epsilon}\frac{d\bar{z}_1}{2\pi i}\bar{z}_1^{n_1}\oint_{|\bar{z}_2|<\epsilon}\frac{d\bar{z}_2}{2\pi i}\bar{z}_2^{n_2}\oint_{|z_{12}|<\epsilon}\frac{dz_1}{2\pi i}\frac{\bar{z}_{12}^{t}}{z_{12}}\bar{\partial}_2^{t-k_1-k_2+4}\gluon^{\tinyzero,c}_{k_1+k_2-1}.
\end{align}
Using the mode expansion for $\bar{\partial}_2^{3-k_1-k_2}\gluon^{\tinyzero}_{k_1+k_2-1}$ and performing the $z_1$ integral, we obtain
\begin{align}
    [\gmode{n_1}{k_1,a},\gmode{n_2}{k_2,b}](z_2)&=\sum_{n\in\mathbb{Z}}\sum_{t=0}^{\infty}\frac{-i f^{abc}}{(t+1)!}\oint_{|\bar{z}_1|<\epsilon}\frac{d\bar{z}_1}{2\pi i}\bar{z}_1^{n_1}\oint_{|\bar{z}_2|<\epsilon}\frac{d\bar{z}_2}{2\pi i}\bar{z}_2^{n_2}\bar{z}_{12}^{t}\bar{\partial}_2^{t+1}\bar{z}_2^{-n-1}\gmode{n}{k_1+k_2-1,c}(z_2).
\end{align}
Carrying out the derivative with respect to $\bar{z}_2$ and summing over $t$ gives
\begin{align}
    [\gmode{n_1}{k_1,a},\gmode{n_2}{k_2,b}](z_2)&=-i f^{abc}\sum_{n\in\mathbb{Z}}\oint_{|\bar{z}_1|<\epsilon}\frac{d\bar{z}_1}{2\pi i}\bar{z}_1^{n_1}\oint_{|\bar{z}_2|<\epsilon}\frac{d\bar{z}_2}{2\pi i}\bar{z}_2^{n_2}\frac{\bar{z}_1^{-n-1}-\bar{z}_2^{-n-1}}{\bar{z}_{12}}\gmode{n}{k_1+k_2-1,c}(z_2).
\end{align}
Evaluating the remaining integrals requires expanding the integrand in $\bar{z}_1$ and $\bar{z}_2$, with the form of the expansion depending on the relative magnitude $|\bar{z}_1|$ compared to $|\bar{z}_2|$. We detail the case $|\bar{z}_1|<|\bar{z}_2|$; the complementary case $|\bar{z}_1|>|\bar{z}_2|$ is analogous and produces the same final result. For $|\bar{z}_1|<|\bar{z}_2|$, the integrand expands as
\begin{align}
    \bar{z}_1^{n_1}\bar{z}_2^{n_2}\frac{\bar{z}_1^{-n-1}-\bar{z}_2^{-n-1}}{\bar{z}_{12}}=\left(\bar{z}_1^{n_1}\bar{z}_2^{n_2-n-2}-\bar{z}_1^{n_1-n-1}\bar{z}_2^{n_2-1}\right)\sum_{i=0}^{\infty}\bar{z}_1^i\bar{z}_2^{-i}.
\end{align}
Performing the $\bar{z}_1$ and $\bar{z}_2$ integrals via the residue theorem then yields
\begin{align}
    [\gmode{n_1}{k_1,a},\gmode{n_2}{k_2,b}](z_2)&=-i f^{abc}[\theta(-n_1-1)-\theta(n_2)]\gmode{n_1+n_2}{k_1+k_2-1,c}(z_2).
\end{align}
If the two step functions take the value of $1$ simultaneously, then we get $(1-1)=0$. Thus only one of the two step functions can take the value of $1$, leading to the commutators
\begin{align}
    [\gmode{n_1}{k_1,a},\gmode{n_2}{k_2,b}](z_2)=
    \begin{cases}
        i f^{abc}\gmode{n_1+n_2}{k_1+k_2-1,c}(z_2)         & \text{for } n_1\geq 2-k_1, n_2\geq 2-k_2,
        \\
        -i f^{abc}\gmode{n_1+n_2}{k_1+k_2-1,c}(z_2)& \text{for } n_1,n_2\leq-1,
    \end{cases}
\end{align}
which closes for $k \leq 1$.
Finally, expanding the holomorphic dependence,
\begin{align}\label{eq:gluon holomorphic mode expansion}
    \gmode{n}{k,a}(z)=\sum_{m\in\mathbb{Z}}z^{-m-1}\gmode{m,n}{k,a},
\end{align}
we obtain the hard current algebra.
\section{Soft current algebra from the hard current algebra}\label{app:soft algebra from hard algebra}

In this Appendix, we demonstrate the validity of the mode expansion
\begin{align}\label{eq:app graviton mode expansion}
    \bar{\partial}^{3-k}\graviton^{\tinyzero}_{k}(z,\bar{z})=\sum_{m,n\in\mathbb{Z}}z^{-m-1}\bar{z}^{-n-1}\hmode{k}{m,n},
\end{align}
employed in the main text, even though $\graviton^{\tinyzero}_{k}$ acts as the logarithmic partner of a primary field. We then reconstruct the commutator $[\gsoftmode{k_1,a}{m_1,-1},\gsoftmode{k_2,b}{m_2,0}]$ from the hard current algebra.

\textbf{Mode expansion.} We define the normalized primaries $\hat{\graviton}^{\Delta,k}(z,\bar{z})\equiv(\Delta-k)\graviton^{\Delta}(z,\bar{z})$. Being primary, they admit the mode expansion
\begin{align}\label{eq:app hath mode expansion}
    \hat{\graviton}_{\Delta,k}(z,\bar{z})=\sum_{m,n\in\mathbb{Z}}z^{-m-1}\bar{z}^{-n}\hhatmode{\Delta,k}{m,n}.
\end{align}
Note that $\hhatmode{k}{m,n}\equiv\hhatmode{\Delta,k}{m,n}|_{\Delta=k}$ are precisely the modes $\hsoftmode{k}{m,n}$ of the conformally soft operator, which are truncated in the anti-holomorphic index to $k-2\leq n\leq 0$. From \eqref{eq:app hath mode expansion}, the subleading operator $\graviton^{\tinyzero}_{k}=\partial_{\Delta}\hat{\graviton}_{\Delta,k}(z,\bar{z})\big|_{\Delta=k}$ expands as
\begin{align}
    &\graviton^{\tinyzero}_{k}(z,\bar{z})=\sum_{m,n\in\mathbb{Z}}z^{-m-1}\bar{z}^{-n}\partial_{\Delta}\hhatmode{\Delta,k}{m,n}\big|_{\Delta=k}.
\end{align}
Applying $\bar{\partial}^{3-k}$ leads to the mode expansion \eqref{eq:app graviton mode expansion}, with the modes related by
\begin{align}\label{eq:graviton hmode=hlogmode}
    \hmode{k}{m,n}=(-n)_{3-k}\partial_{\Delta}\hhatmode{\Delta,k}{m,n-2+k}\big|_{\Delta=k}.
\end{align}

\textbf{More examples of reconstruction.} In the main text we derived
\begin{align}\label{eq:app ghat0 ghat0}
    [\gsoftmode{k_1,a}{m_1,0},\gsoftmode{k_2,b}{m_2,0}]=-if^{abc}\frac{(2-k_1-k_2)!}{(1-k_1)!(1-k_2)!}\gsoftmode{k_1+k_2-1,c}{m_1+m_2,0}.
\end{align}
We now reconstruct additional commutators. Consider first $[\gsoftmode{k_1,a}{s_1,-1},\gsoftmode{k_2,b}{s_2,0}]$. Acting with $\ad^{3-k_1-k_2}_{\bar{L}_1}$ on $[\gmode{s_1,-1}{k_1,a},\gmode{s_2,-1}{k_2,b}]$ and with $\ad_{\bar{L}_{-1}}$ on \eqref{eq:app ghat0 ghat0} gives two independent equations:
\begin{align}
    &[\gsoftmode{k_1,a}{s_1,-1},\gsoftmode{k_2,b}{s_2,0}]+[\gsoftmode{k_1,a}{s_1,0},\gsoftmode{k_2,b}{s_2,-1}]=-if^{abc}\frac{(2-k_1-k_2)!}{(1-k_1)!(1-k_2)!}\gsoftmode{k_1+k_2-1,c}{s_1+s_2,-1},\\\nonumber
    &C1[\gsoftmode{k_1,a}{s_1,0},\gsoftmode{k_2,b}{s_2,-1}]+C_2[\gsoftmode{k_1,a}{s_1,-1},\gsoftmode{k_2,b}{s_2,0}]=-if^{abc}\frac{(4-k_1-k_2)!}{(2-k_1-k_2)_2}\gsoftmode{k_1+k_2-1,c}{s_1+s_2,-1},
\end{align}
where $C_1 = (2-k_2)k_2!(1-k_1)!$, and $C_2 = (2-k_1)k_1!(1-k_2)!$. Solving these equations yields
\begin{align}\label{eq:app ghat1 ghat0}
    [\gsoftmode{k_1,a}{s_1,-1},\gsoftmode{k_2,b}{s_2,0}]=-if^{abc}\frac{(1-k_1-k_2)!}{(-k_1)!(1-k_2)!}\gsoftmode{k_1+k_2-1,c}{s_1+s_2,-1},
\end{align}
which matches the soft current algebra in \cite{Guevara:2021abz}.


\clearpage
\printbibliography


\end{document}